\documentclass[reprint,aps,prl,amsmath,longbibliography,superscriptaddress]{revtex4-1}
\usepackage{color}
\usepackage{graphicx}
\usepackage{dcolumn}
\usepackage{bm}
\usepackage{float}
\usepackage{siunitx}
\usepackage[normalem]{ulem}

\usepackage[colorlinks=true,
  urlcolor=blue,
  linkcolor=blue,
  citecolor=blue]{hyperref}
  
\newcommand{\normord}[1]{\vcentcolon\mathrel{#1}\vcentcolon}
\providecommand{\vcentcolon}{\mathrel{\mathop{:}}}

\newcommand{\ctwo}{C^{(2)}(d)}
\newcommand{\cthree}{C^{(3)}(d_1,d_2)}
\newcommand{\nknk}{\langle \hn_{k1} \hn_{k2} \rangle}
\newcommand{\nknknorm}{\langle \normord{\hn_{k1} \hn_{k2}} \rangle}

\newcommand{\nknknk}{\langle \hn_{k_1} \hn_{k_2} \hn_{k_3} \rangle}
\newcommand{\nknknknorm}{\langle \normord{\hn_{k_1} \hn_{k_2} \hn_{k_3}} \rangle}

\newcommand{\klattice}{k_\textrm{lat}}
\newcommand{\diff}{\mathop{}\!\mathrm{d}}
\newcommand{\lend}{\nonumber \\}
\newcommand{\hn}{\hat{n}}
\newcommand{\hkn}[1]{\hat{n}_{k_{#1}}}
\newcommand{\dn}[1]{\langle \hat{n}_{k_{#1}} \rangle}

\newcommand{\beginsupplement}{%
        \setcounter{table}{0}
        \renewcommand{\thetable}{S\arabic{table}}%
        \setcounter{figure}{0}
        \renewcommand{\thefigure}{S\arabic{figure}}%
     }

\begin{document}
\title{High-Contrast Interference of Ultracold Fermions}
\author{Philipp M. Preiss}
\email{preiss@physi.uni-heidelberg.de}
\affiliation{Physics Institute, Heidelberg University, 69120 Heidelberg, Germany}
\author{Jan Hendrik Becher}
\affiliation{Physics Institute, Heidelberg University, 69120 Heidelberg, Germany}
\author{Ralf Klemt}
\affiliation{Physics Institute, Heidelberg University, 69120 Heidelberg, Germany}
\author{Vincent Klinkhamer}
\affiliation{Physics Institute, Heidelberg University, 69120 Heidelberg, Germany}
\author{Andrea Bergschneider}
\thanks{Present address: Institute for Quantum Electronics, 8093 Zurich, Switzerland}
\affiliation{Physics Institute, Heidelberg University, 69120 Heidelberg, Germany}
\author{Nicol\`{o} Defenu}
\affiliation{Physics Institute, Heidelberg University, 69120 Heidelberg, Germany}
\affiliation{Institute for Theoretical Physics, Heidelberg University, 69120 Heidelberg, Germany}
\author{Selim Jochim}%
\affiliation{Physics Institute, Heidelberg University, 69120 Heidelberg, Germany}%

\date{\today}      
\begin{abstract}
Many-body interference between indistinguishable particles can give rise to strong correlations rooted in quantum statistics. We study such Hanbury Brown-Twiss-type correlations for number states of ultracold massive fermions. Using deterministically prepared $^6$Li atoms in optical tweezers, we measure momentum correlations using a single-atom sensitive time-of-flight imaging scheme. The experiment combines on-demand state preparation of highly indistinguishable particles with high-fidelity detection, giving access to two- and three-body correlations in fields of fixed fermionic particle number. We find that pairs of atoms interfere with a contrast close to 80\%.  We show that second-order density correlations arise from contributions from all two-particle pairs and detect intrinsic third-order correlations.

\end{abstract}

\maketitle

Many-body interference describes processes by which non-interacting particles acquire strong correlations solely due to their quantum statistics \cite{Tichy2014}. In such cases, interference occurs between many-particle paths and the enhancement or suppression of particular outcomes is dictated by the exchange statistics of the particles.

The most famous example of many-body interference is the ``bosonic bunching" of photons, as observed by Hanbury Brown and Twiss in the development of stellar intensity interferometry \cite{Brown1956,Brown1956b}. The experimental \cite{Ghosh1987, Hong1987} and theoretical \cite{Glauber1963} study of correlations in photon fields has been driving the development of quantum optics \cite{Scully1997}.

In contrast to the statistical behavior of bosons, which can in certain cases be described in terms of classical waves \cite{Brown1956}, the interference of fermionic particles is a uniquely quantum mechanical phenomenon. Experimental access to correlations arising from fermionic interference allows the study of quantum systems through intensity interferometry, for example in heavy ion collisions \cite{Alexander2003} and ultracold atom systems \cite{Rom2006}. Fermionic interference and  associated anti-correlations have been observed for thermal sources of neutrons \cite{Iannuzzi2006} and cold atoms \cite{Jeltes2007}, as well as electrons in free space \cite{Kiesel2002} and in solid state \cite{Oliver1999,Henny1999}. Interference between fermionic number states, however, has been far more elusive: Few particle species amenable to single-particle control possess fermionic statistics, and realizing high-quality single-fermion sources is an outstanding experimental challenge. Recent advances in this direction have enabled the observation of two-fermion interference in semiconductor architectures \cite{Bocquillon2013} and double ionization processes \cite{Waitz2016}. 
\begin{figure}[ht!]
	\centering
	\includegraphics{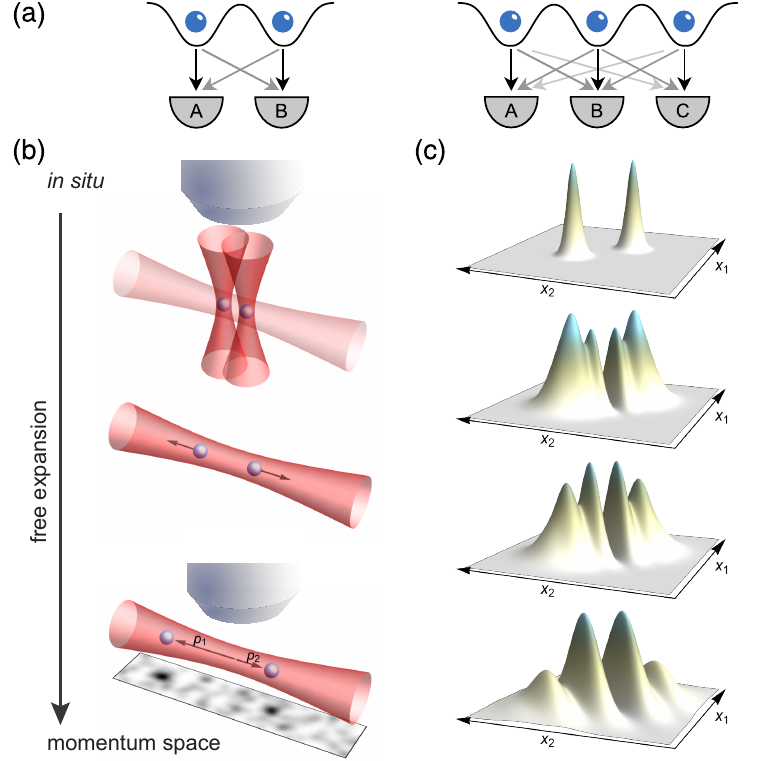}
	\caption{Many-body interference of fermionic particles. (a) The coincidence events for two detectors monitoring single-particle sources are given by the available sets of many-body paths. For indistinguishable particles, the paths add coherently, where the sign of their interference depends on quantum statistics. Two sources emitting indistinguishable fermions display suppressed coincidence counts. In a setting with more particles, correlations between detectors arise due to interference of all many-body paths. (b) We realize the thought experiment from (a) using ultracold fermions in optical tweezers. Time-of-flight expansion is performed in an optical dipole trap aligned with the $x$-axis connecting the tweezers and particles are detected in free-space fluorescence imaging. (c) The probability distribution $|\Psi(x_1,x_2)|^2$ (plotted for the two-mode case) evolves from localized to delocalized states during time-of-flight, but retains a node at $x_1=x_2$ due to fermionic anti-symmetry.}
	\label{fig1}
\end{figure}

Here, we observe high-contrast interference of fermionic particles in pure quantum states of ultracold atoms. We use optical tweezers as a configurable, deterministic source of non-interacting fermions \cite{Serwane2011, Murmann2015}. The particles interfere during a ballistic time-of-flight and are detected with high fidelity. We observe periodic anti-bunching of two fermions with close to full contrast.  Adding a third source to the system, we measure the second- and third-order correlation properties of a three-fermion field and determine the contribution of intrinsic third-order correlations \cite{Cantrell1970,Liu2016}.
 \begin{figure}
	\centering
	\includegraphics{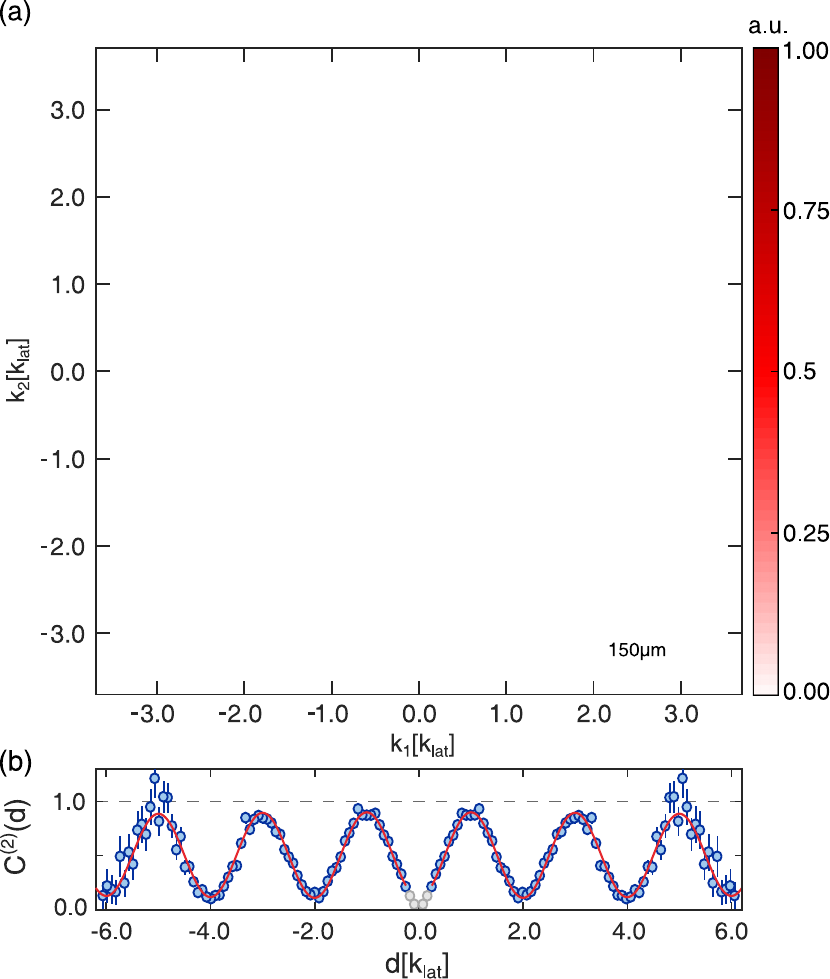}
	\caption{Interference of two fermions.  a) Measured momentum correlator $\nknknorm$ for two particles prepared in two optical tweezers, showing strong fermionic anti-bunching. The data are symmetric about the diagonal $k_1 = k_2$ by construction, but we plot the full correlator for visual clarity. 
	(b) The normalized correlator $\ctwo$ shows a clear, almost full sinusoidal modulation in the relative momentum $d$. We extract a contrast of $79(2)\%$ by fitting a cosine to the data excluding the greyed out points near the origin, for which coincidences cannot reliably be detected.}
	\label{fig2}
\end{figure}

Our experiments use $^6$Li atoms in the same internal state initialized in optical microtraps separated by tunable distances of several microns (see Fig.\,\ref{fig1})\cite{Methods}. Within the regime of $s$-wave interactions, fermions in the same internal state do not interact, and the particles behave as ideal free fermions. We release the atoms simultaneously into a weak, elongated optical dipole trap aligned with the axis connecting the tweezers. The dipole trap enables a one-dimensional time-of-flight measurement, wherein the wavepackets remain localized in the transverse directions. In the weakly confined $x$-direction, evolution in the approximately harmonic confining potential linearly maps initial momenta onto final position \cite{Murthy2014}, similar to free-space time-of-flight schemes \cite{Foelling2005, Rom2006}. The single-particle wavefunctions are initially localized to $\sim$\SI{250}{\nano \metre} and expand to a size of $\sim$\SI{400}{\micro \metre} before detection. We extract the position of individual atoms along the axis of the dipole trap with a free-space state-resolved imaging technique \cite{Bergschneider2018} (see Fig.\,\ref{fig1}\,(b)). Any correlations can be interpreted as position correlations upon interference or equivalently as momentum correlations in the initial state.

In a first set of experiments we study the interference of fermionic particles from two tweezers separated by a distance of $a=\SI{1.7}{\micro \metre}$. The tweezers are loaded independently from a degenerate Fermi gas and initialized with the $n=1$ number state with a probability of $97(2)$\% per tweezer. Residual tunnel couplings of at most \SI{0.1}{Hz} are negligible on the $\sim$\SI{100}{\milli \second} time scale of the experiment. This configuration corresponds to the scenario sketched in Fig.\,\ref{fig1}\,(a) and (c): Fermionic correlations arise due to destructive interference of two sets of two-particle paths, or equivalently from the preservation of anti-symmetry of the two-particle wavefunction during time-of-flight. This experiment directly realizes the fermionic analog of the Ghosh-Mandel experiment \cite{Ghosh1987}. 

\begin{figure*}[ht]
	\centering
	\includegraphics{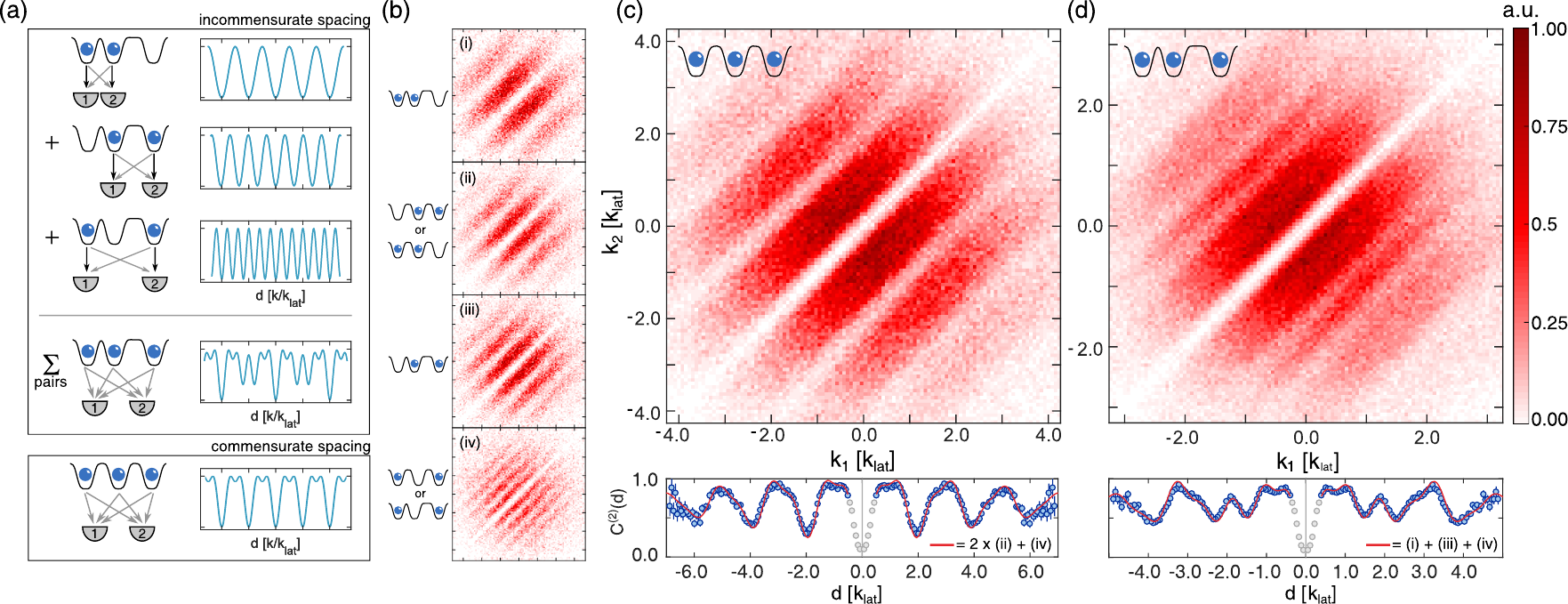}
	\caption{Second-order correlations for three fermions. (a) Top: For an arbitrary configuration of single-fermion sources, each pair of particles contributes a sinusoidal oscillation to the correlation function $\ctwo$. The full correlator can exhibit complex patterns due to beat notes between all spatial frequencies. Bottom: Expected full correlation function for three regularly spaced sources. (b) Correlators for individual pairs of sources. (c) Three particles released from three equidistant tweezers exhibit correlations at the momentum scale $k_\textrm{lat}$ and its first harmonic, leading to a ``superlattice" structure in $\ctwo$. The bottom row displays the measured $\ctwo$ for three particles (data points), together with a prediction (red line) given by the weighted sum of fits to the two-particle contributions from panel b). (d) For three tweezers with irregular spacing, three distinct spatial frequencies contribute to the full second-order correlator. The sharp dip at $k_1 = k_2$ is partially due to our lack of sensitivity to coincidences at short distances.}
	\label{fig3}
\end{figure*}

Our method extends noise-correlation analysis, which was pioneered on many-body lattice systems \cite{Altman2004, Foelling2005, Rom2006, Spielman2007}, to measurements of full atomic correlations of a single, nearly pure quantum state. The availability of individual particle momenta allows us to obtain normal-ordered correlation functions that are free from autocorrelation peaks at zero interparticle distance \cite{FoellingBook2015, Methods}.  We record the momenta of the two particles for each experimental run and construct the momentum density  $\langle n_k \rangle$ at wavevector $k$ and its second-order correlator $\nknknorm$ from several thousand realizations ($\langle \normord{\cdot}  \rangle$ denotes normal ordering). We only retain data from runs where two particles were successfully prepared and detected, which corresponds to about $80\%$ of the data.

In Fig.\,\ref{fig2}\,(a) the experimentally measured momentum correlator is shown. The strong correlations in the relative momentum $d=k_1-k_2$ immediately demonstrate the non-separability of the correlator into single-particle momenta. The modulations occur on a ``lattice" momentum scale $\klattice=\pi/a$ and carry an envelope set by the single-particle on-site momentum distribution.   

As correlations do not depend on the centre of mass momentum, we define a normalized correlation function \cite{Foelling2005, Rom2006}
\begin{equation}
\label{correlator}
    C^{(2)}(d) =  \frac{\int\langle \normord{ \hn_{k} \hn_{k+d}}\rangle \diff k}{\int\langle \hn_{k}\rangle \langle \hn_{k+d} \rangle \diff k}.
\end{equation}
The correlation function in Fig.\,\ref{fig2}\,(b) exhibits close to full-contrast sinusoidal oscillations consistent with a minimum at $d=0$, corresponding to strong fermionic anti-bunching. The greyed out points near $d=0$ mark the region of particle separation below \SI{30}{\micro \metre}, where two particles cannot be distinguished reliably and coincidences cannot be detected in our experiment \cite{Methods}. We exclude these points from all further analysis. To quantify the strength of periodic correlations, we fit a damped cosine function to the correlations away from $d=0$, which gives a modulation contrast of $79(2)\%$. The contrast certifies the high degree of indistinguishability between the particles during the preparation and the matterwave expansion. With a ground state preparation fidelity of $97(2)\%$, one would in principle expect correlations with a contrast of up to  $\sim$95$\%$. The measured contrast is limited by alignment errors between the axis of the optical dipole trap and the tweezer axis, which provide some distinguishability between the particles during the expansion dynamics \cite{Methods}. Nevertheless, the contrast significantly exceeds the minimum visibility of  $1/\sqrt{2}\approx 71\%$ required to perform quantum optics experiments such as non-locality tests with massive particles \cite{Samuelsson2004, Bonneau2018}.

\begin{figure*}[ht]
	\centering
	\includegraphics{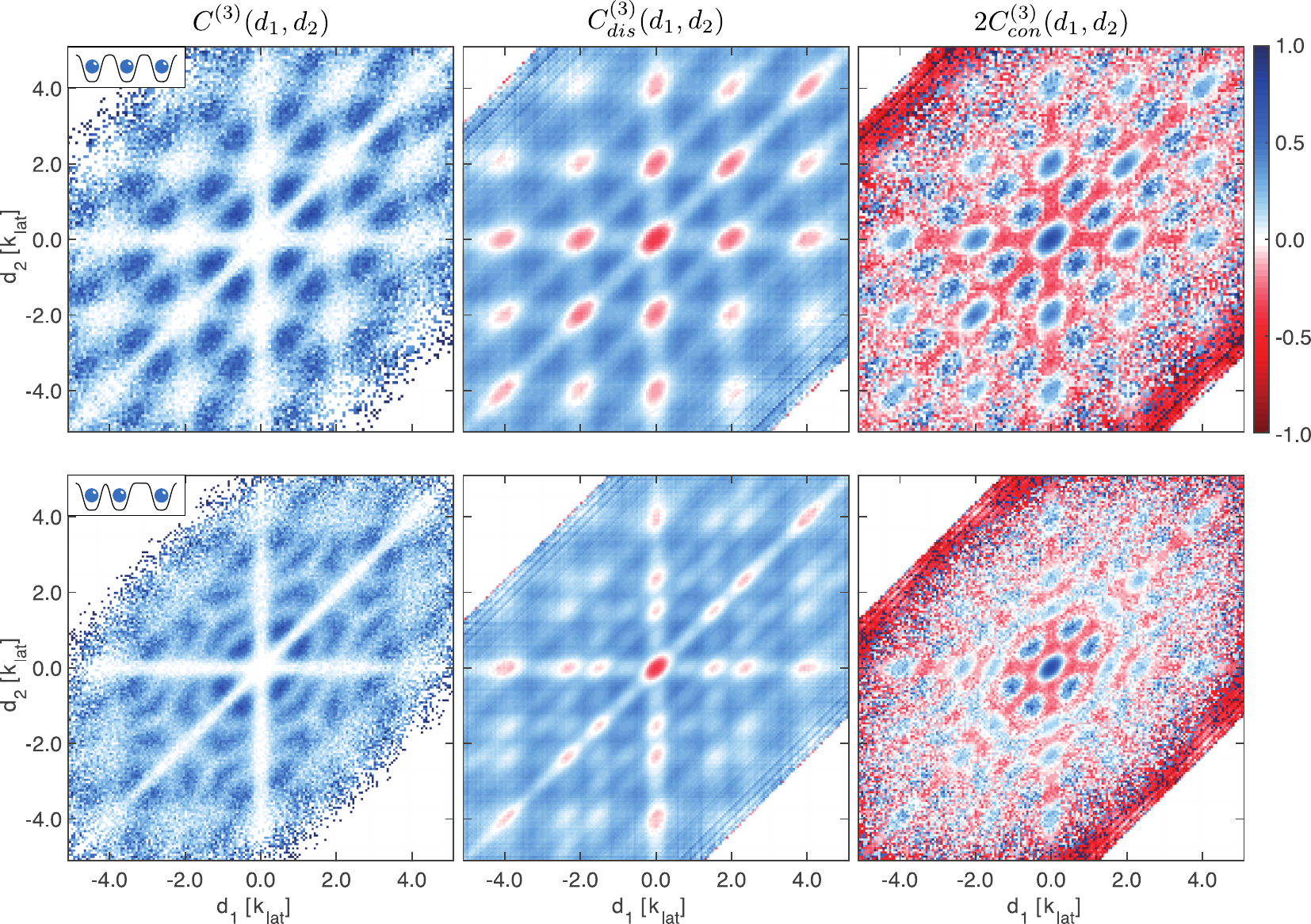}
	\caption{Third-order momentum correlations. Normalized correlation function $C^{(3)}(d_1,d_2)$ for three fermions released from three wells with regular spacing ($a_{23}=a_{12}$, top row) and non-equal spacing ($a_{23}=1.5 a_{12} $, bottom row). The left column shows the full measured correlations, the central and right column show the disconnected and connected part of the correlation function, respectively. By construction, the function is symmetric under the transformation $(d_1, d_2) \Leftrightarrow (-d_1, -d_2)$. The connected part of the correlation function $C^{(3)}_{con}$ is scaled by a factor of two for better visibility. The measurement demonstrates the sensitivity to third-order correlations in the fermionic matterwave field.}
	\label{fig4}
\end{figure*}

With this high-contrast, on-demand source of indistinguishable fermions, we can now study second- and third-order correlations of a triplet of sources (Fig.\,\ref{fig1}). For a system of $N$ localized sources each emitting a single fermionic particle, it can be shown that negative exchange symmetry gives rise to a second-order correlation function of the form
\begin{equation}
C^{(2)}(d)  =  \frac{2}{N^2} \sum_{<i,j>}^N(1 - \cos{(d (x_i - x_j))},
\label{eq:ctwo}
\end{equation}
where summation runs over all distinct pairs of emitters and $x_i$ refers to the center of the $i^\textrm{th}$ source \cite{Methods}. 

This expression gives an intuitive picture for the expected second-order momentum correlations (Fig.\,\ref{fig3}\,(a)): Every pair of particles generates a sinusoidal anti-bunching signal with a spacing given by the inverse source separation. The full second-order correlation function is the sum over all pairwise correlation signals. 

We probe the validity of this picture using three equidistant tweezers with $a_{12}=a_{23}=\SI{2}{\micro \metre}$, where $a_{ij}$ refers to the spatial separation of tweezer $i$ and $j$. Pairs from neighbouring tweezers contribute a `long wavelength' correlation signal at frequency $k_{lat} = \pi/a_{12} = \pi/a_{23}$, whereas the outermost tweezers 1 and 3 give rise to a `short wavelength' modulation with half the period $\pi/a_{13}=\klattice/2$. The two contributions should result in a ``superlattice" correlation structure, shown in Fig.\,\ref{fig3}\,(a). Adding more sources, and hence Fourier components, would lead to narrower correlation minima and result in the delta-function correlations observed for fermionic band insulators \cite{Rom2006}.

Figure\,\ref{fig3}\,(b) shows the experimentally measured second-order correlations. We first record the correlator for every tweezer spacing individually by loading only two of the three microtraps. The pairwise correlations (shown in panels (i) to (iv)) are identical to the two-tweezer case in Fig.\,\ref{fig2}, with a correlation scale given by the inverse spacing of the active sources. Separately we measure the full second-order correlator for all three tweezers loaded simultaneously (Fig.\,\ref{fig3}\,(c)).
 
The bottom panels of Fig.\,\ref{fig3}\,(c) and (d) show the correlation function $\ctwo$ for the three particle system (blue data points). We test the applicability of Eq.(\ref{eq:ctwo}) by a comparison to the two-tweezer contributions: We individually fit the correlation functions from panel (b) and show the weighted sum of the fits as the red line \cite{Methods}. The three-particle second-order correlation function, including the decay of the contrast towards larger relative momenta $d$, is very well reproduced by the sum of the pair contributions. 

For the above case of regularly spaced sources, all two-particle correlations occur at the lattice momentum scale and its first harmonic. Source arrays with irregular spacing, on the other hand, can result in distinct correlation signals from all pairs of emitters and complex structures in the full correlator (see Fig.\,\ref{fig3}\,(a)).

We study the case of three tweezers with non-equal spacing in Figure\,\ref{fig3}\,(d), where $a_{12} =\SI{1.6}{\micro \metre}$, $a_{23}=1.5\,a_{12} =\SI{2.4}{\micro \metre}$. The incommensurate spacing leads to a doubling of the unit cell to four times the momentum $\klattice$, which we define via the smallest tweezer spacing $a_{12}$. Also here the correlations in the three-particle system are in perfect agreement with the weighted contributions from pairs of sources.\\

To fully characterize the field produced by multiple sources, a measurement of correlation functions at higher orders is required. For bosonic particles, such measurements are routinely performed to assess the statistical properties of light sources \cite{Scully1997,Assmann2009,Singer2013}. Recently, measurements of higher-order correlations of massive bosonic particles have become possible with ultracold atoms \cite{Dall2013, Schweigler2017,Hodgman2017}.

We measure the statistical properties of the matterwave field emanating from the three fermionic sources via the third-order correlator $\nknknknorm$ and the corresponding normalized correlation function
\begin{equation}
    C^{(3)}(d_1,d_2) = \frac{\int \langle \normord{ \hn_{k} \hn_{k+d_1} \hn_{k+d_2}}\rangle \diff k}{\int\langle \hn_{k}\rangle \langle \hn_{k+d_1} \rangle \langle \hn_{k+d_2} \rangle \diff k}.
\end{equation}

The correlation function is shown in Fig.\,\ref{fig4} for the two data sets from Fig.\,\ref{fig3}. The equidistant and incommensurate tweezer configurations lead to clear and distinct correlation features at third order. 

In order to interpret the third-order density correlations, it is useful to first remove contributions from lower order.  This can be achieved by subtracting a suitable combination of first- and second-order correlators from the full third-order correlation function. Any remaining correlations are intrinsic, that is they cannot be accessed from measurements at lower order. In interacting systems, such intrinsic correlations carry crucial information about the many-body state \cite{Schweigler2017,Hodgman2017}, but they may be present even for free particle systems \cite{Cantrell1970, Liu2016}. For non-interacting bosons, for example, the intrinsic correlations contribute to a striking increase in zero-distance correlations at higher order \cite{Assmann2009,Dall2013}. 
 
To assess the presence of intrinsic third-order correlations in our system, we combine our measurements at second order to construct the disconnected part of the third-order correlation function $C^{(3)}_{dis}(d_1,d_2)$. We define the disconnected correlator as \cite{Cantrell1970,Liu2016,Schweigler2017}
 
\begin{align}
    &\nknknknorm_{dis} = \lend
 & s_1(N) \left (\langle \hn_{k_1} \rangle \langle \normord{\hn_{k_2} \hn_{k_3}}\rangle + \langle \hn_{k_2} \rangle \langle \normord{\hn_{k_1} \hn_{k_3}}\rangle + \langle \hn_{k_3} \rangle \langle \normord{ \hn_{k_1} \hn_{k_2}}\rangle \right ) \lend
     & \quad \quad - 2 s_2(N) \langle \hn_{k_1} \rangle \langle \hn_{k_2} \rangle \langle \hn_{k_3}\rangle.
     \label{eqn:discon}
\end{align}

The scale factors $s_1(N) = \frac{N(N-1)(N-2)}{N^2(N-1)}$ and $s_2(N) = \frac{N(N-1)(N-2)}{N^3}$ account for correlations due to particle number conservation \cite{Methods} and approach unity as $N\rightarrow \infty$. The corresponding correlation function $ C^{(3)}_{dis}(d_1,d_2)$ represents the experimentally accessible knowledge of third-order correlations available from pairwise measurements and is shown in the central column of Fig.\,\ref{fig4}. Clearly, it does not include all features of the full correlation function and the connected part of the correlation function, $C^{(3)}_{con} = C^{(3)} - C^{(3)}_{dis}$, retains additional structure, shown in the right hand column of Fig.\,\ref{fig4}. The presence of intrinsic correlations agrees very well with an analytic calculation and is in full agreement with a decomposition according to Wick's theorem \cite{Wick1950, Methods}. Recovering the expected functional form of third-order correlations for different spatial tweezer arrangements validates our analysis. We conclude that the system displays strong correlations at third order consistent with ideal fermionic statistics. To our knowledge, this constitutes the first experimental characterization of a fermionic field beyond second order and realizes the fermionic counterpart to the recent observation of three-photon interference \cite{Agne2017, Menssen2017}.

Our work opens the door for several interesting avenues of research: Our high-purity, on-demand source of indistinguishable fermions may enable quantum optics experiments with massive particles, such as fermionic ghost imaging or Bell tests \cite{Liu2016, Khakimov2016, Bonneau2018}. Extending our methods to more particles and modes, the interplay of coherence, indistinguishability and quantum statistics can be studied in many-fermion interference \cite{Tichy2014}. This is of particular interest in regards to dynamics of many-body systems, which are given by a combination of interaction effects and the interference phenomena studied here  \cite{Brunner2018}.

\begin{acknowledgments}
We thank M. G\"{a}rttner, P. Hauke, C. Westbrook and G. Z\"{u}rn for insightful discussions. This work has been supported by the ERC consolidator grant 725636, the DFG grant JO970/1-1, the Heidelberg Center for Quantum Dynamics, and the DFG Collaborative Research Centre SFB 1225 (ISOQUANT). A. B. acknowledges funding from the International Max-Planck Research School (IMPRS-QD). P.M.P. acknowledges funding from EU Horizon 2020 programme under the Marie Sklodowska-Curie grant agreement No. 706487 and the Daimler and Benz Foundation.
\end{acknowledgments}



%


\clearpage
\beginsupplement
\section{Methods}
\subsection{\label{System}Experimental Procedure}
\subsubsection{State Preparation and Imaging}
We prepare $^6$Li atoms in optical tweezer traps following the procedure outlined in references \cite{Serwane2011, Murmann2015}: Using an accusto-optical deflector (AOD) and trapping light at wavelength $\lambda=\SI{1064}{\nano \metre}$ we create several tweezer traps with a waist of $w=\SI{1.15}{\micro \metre}$ and tunable spatial displacement \cite{Murmann2015}. We independently prepare particles in the ground state of each tweezer, ensuring that tunneling between the wells is fully suppressed at all times. 

Time-of-flight (TOF) is performed in a weak elongated optical dipole trap (ODT) aligned along the axis connecting the tweezers \cite{Bergschneider2018} (see next section). For all experiments we extract the 1D momentum of the atoms along the $x$-axis of the ODT by measuring the position after TOF in the ODT with a single-atom state-resolved imaging technique \cite{Bergschneider2018}. The imaging system has an effective resolution of \SI{4}{\micro \metre} and the distance between two atoms of the same hyperfine state beyond which they can be distinguished with more than $90\%$ fidelity is $\sim$\SI{30}{\micro \metre}. This minimal spacing sets a ``dead volume" for the detectors, within which coincidences cannot be detected.
We independently tune the trap depth of the tweezers and therefore the size of the on-site wave function, the separation of the tweezers, and the trap frequency of the confining potential during TOF. We choose parameters to compromise between the single-particle envelope, correlation length scales, imaging resolution and imaging dead volume. 

For our experiments, we prepare two particles per well, one each in two of the three lowest hyperfine states, commonly labelled $|1\rangle$ and $|3\rangle$, ascending in energy. Using a magnetic Feshbach resonance we tune the $s$-wave scattering length between the two hyperfine states to a zero crossing at \SI{568}{G}, such that there are no interactions between them. We take separate images of both hyperfine states during each experimental run \cite{Bergschneider2018} and consider the two spin manifolds as two independent realizations of the same setting, effectively increasing the statistics by a factor of two. We have verified that there are no correlations between the two different spin manifolds. Preparation fidelities are $\sim95\%$ for two atoms in a single tweezer and correspondingly $\sim 85\%$ for a six-particle system.

\subsubsection{\label{System}Time-of-Flight}

\begin{figure}
    \centering
    \includegraphics{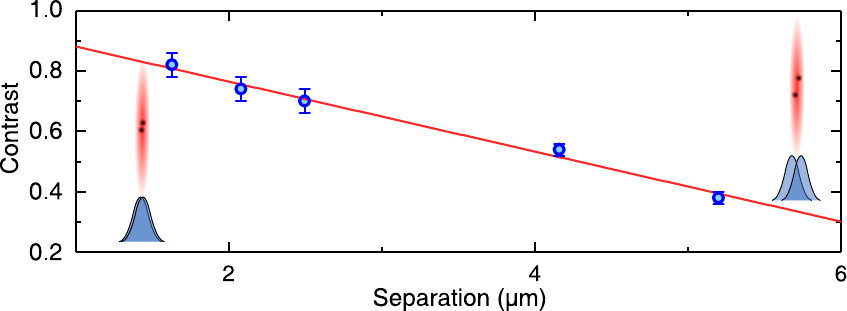}
    \caption{The measured contrast of two-fermion interference, measured via $C^{(2)}(d)$, decreases for increasing separation between the tweezers. We attribute this effect to the residual angle between tweezer axis and cODT, which leads to a reduced initial mode overlap of the tweezers in the frame defined by the cODT axes (insets). The red line is a linear fit to the data.}
    \label{separation}
\end{figure}

All experiments except the measurement for Fig.\,2 of the main text are performed in a crossed-beam optical dipole trap (cODT). The trap frequencies during TOF in the cODT are $( \omega_x,\omega_y,\omega_z ) \approx 2\pi\times ( 40, 320, 320 )$\,Hz and time of flight lasts \SI{7}{\milli \second}. Figure\,\ref{separation} shows a summary of the measured contrasts for two-particle interference experiments with different well separations. The contrast decreases for increasing separation between the tweezers. We attribute this effect to the residual angle of $1^\circ$ between the cODT and the tweezer axis. As shown in the insets in Fig.\,\ref{separation}, a large distance between the tweezers corresponds to a small mode overlap of the wave functions along the axis of the cODT. This effect is particularly relevant for large separation between the wells, i.e. for small momentum scales and currently sets the main limitation on the measured contrast. 

The expansion for Fig.\,2 of the main text was performed in a single beam ODT with only a very small harmonic confinement along the long axis. The residual angle between ODT and the tweezer axis in this case is $\sim5^\circ$. 

\subsection{\label{System}Data Processing}

For each experimental setting we record several thousand realizations and postselect all recorded images for the correct atom number. We achieve a postselection rate of $\sim60\%$ for the three-atom setting and $\sim80\%$ for the two-atom experiments. The difference between the preparation fidelity and the postselection rate is caused primarily by atoms in the high-$k$ wings of the distributions, which we neglect, and by the fact that atoms in close proximity to each other cannot be detected.

Prior to analysis, we group the measured positions after TOF into bins defined by two to four camera pixels.

\subsubsection{Fitting the Correlator}
In order to extract the contrast of the measured correlations we fit a phenomenological model to the correlation functions. For experiments with two fermions we fit a model with a single frequency
\begin{multline}
    C^{(2)}(d) = \\
    \frac{1}{2}\left(1+\text{erf}\left(\frac{|d|-s}{w}\right)\right)\left(y_0- e^{-\frac{d^2}{2\xi^2}}c\cos{(\pi d/ \klattice)}\right).
    \label{fit}
\end{multline}

The error function parameterized by $s$ and $w$ accounts for the dead volume of the imaging system. The exponential length scale $\xi$ accounts for the decay of the contrast $c$ for large separations $d$. $k_{lat}$ is the lattice momentum and $y_0$ the offset. The free parameters of the fit are $s$, $w$, $y_0$, $\xi$, $c$, and $\klattice$.

\subsubsection{Quadratic Corrections of Single Particle Momenta} 
\begin{figure}
    \centering
    \includegraphics{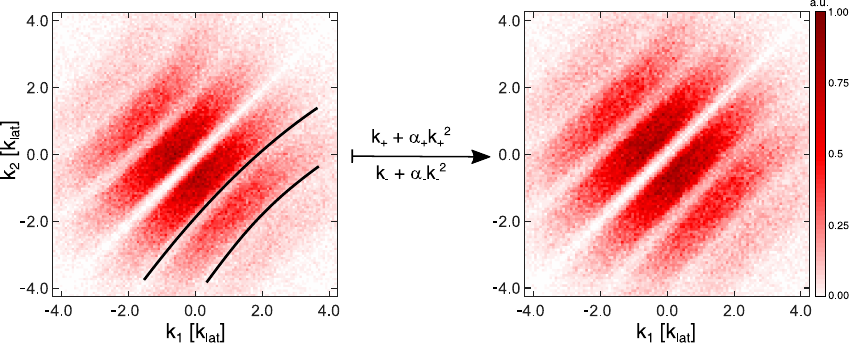}
    \caption{We remove distortions of the momentum correlation function by quadratically rescaling the single particle momenta. The two coefficients $\alpha_\pm$ are chosen to optimize the contrast and the decay length $\xi$ of the mometum correlator $C^{(2)}(d)$. The procedure is shown here for the data presented in Fig.\,3 (c) of the main text.}
    \label{quadr}
\end{figure}
For an expansion for a quarter of a trap period in a  harmonic trap, there is an exact linear relation $k(x) = \tilde{k} x$ between initial momentum $k$ and position $x$ after TOF \cite{Murthy2014}. However, the geometry of the ODT is not strictly harmonic but rather Gaussian and as we expand the atomic wave function to roughly the waist of the trap, anharmonic corrections start to play a role. As a consequence the correlator is distorted, influencing the contrast $c$ and decay length $\xi$ in Eq.\,(\ref{fit}), see left panel in Fig.\,\ref{quadr}. We correct this effect by taking a quadratic correction to the mapping $k(x)$ into account. We rescale the momentum of each particle as $k_\pm \mapsto k_\pm + \alpha_\pm k_\pm^2$ and choose $\alpha_\pm$ to maximize $c$ and $\xi$ of the fit to the correlation function. To account for the asymmetric distortion of the correlator, we use two rescaling parameters, $\alpha_+$ for positive and $\alpha_-$ for negative momenta. This rescaling removes the distortions from the correlator, see right panel in Fig.\,\ref{quadr}. For momenta measured in units of $k_{lat}$, $\alpha_\pm$ takes typical values of $\sim 0.025$. The observed quadratic corrections are in reasonable agreement with a numerical estimate of the effect of the Gaussian trap shape. For a $1/e^2$ trap size of $L_\textrm{ODT}=\SI{400}{\micro \metre}$ and a $\klattice$-equivalent displacement of $\SI{74}{\micro \metre}$, we would expect a rescaling coefficient $\alpha = 0.083$.

\subsubsection{Reconstruction of the Second-Order Correlation Function} 
\begin{figure*}
    \centering
    \includegraphics{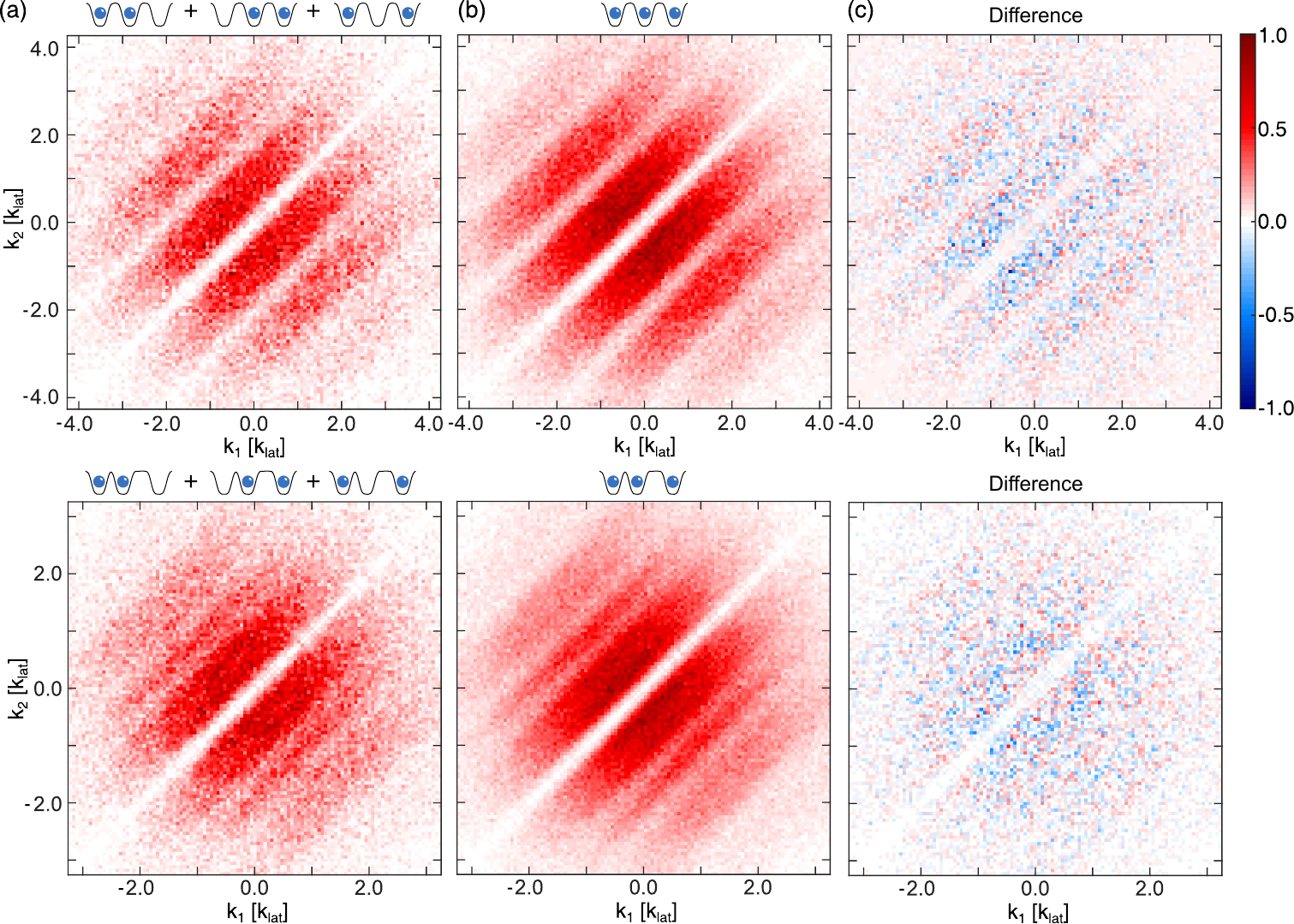}
    \caption{Reconstruction of the second-order three-atom correlator. (a) shows the mean of all mutual measured two-atom correlators. (b) shows the measured second-order correlator for three fermions released from three tweezers. (c) shows the difference (b)-(a). There is almost no structure in (c), verifying that the full second-order correlator can be reconstructed from the measured two-atom correlators.}
    \label{reconstruction}
\end{figure*}
The second-order correlation function for three atoms released from three wells is a sum over all mutual two-particle correlation functions (see Eq.\,(2) of the main text). The two-particle correlator for every tweezer spacing was measured in the experiment and is shown in Fig.\,3 of the main text. Figure\,\ref{reconstruction} (a) shows the reconstructed correlator, i.e. the sum of the three two-atom correlators for the two different well configurations. The measured three-atom correlator is shown in (b). Note that the dataset for the latter is much larger than those for the two-atom correlator and therefore the noise is lower in (b). Figure\,\ref{reconstruction} (c) shows the difference of (a) and (b). There is almost no structure left, verifying that the full second-order correlator can be reconstructed from the individual two-atom correlators,
\begin{align}
    \langle \hn_{k_1} \hn_{k_2}\rangle_{123} = &\langle \hn_{k_1} \hn_{k_2}\rangle_{12} \lend 
    +&\langle \hn_{k_1} \hn_{k_2}\rangle_{23} \lend
    +&\langle \hn_{k_1} \hn_{k_2}\rangle_{13}.
\end{align}

Here, $\langle \hn_{k_1} \hn_{k_2}\rangle_{ijk}$ denotes the correlator for the initial state with atoms loaded into tweezers $i$, $j$, $k$.

\subsection{Theoretical Analysis}

\subsubsection{Comparison to Noise Correlations Measurements}

The method we employ here is closely related to noise correlation measurements \cite{Altman2004} performed on many-body lattice systems of bosons \cite{Foelling2005, Spielman2007} and fermions \cite{Rom2006}. There are several technical differences to typical density-density correlation measurements performed with absorption imaging: (i) The spatial scale of correlation features after time-of-flight is much larger than the imaging resolution and we directly probe the atomic correlations without additional broadening or averaging mechanisms; (ii) we avoid line-of-sight integration of correlation features by working in a one-dimensional geometry; (iii) we prepare a single, well-defined quantum state, avoiding an inhomogeneous average of the correlation function over regions with different filling.

The most important conceptual difference to noise correlation measurements is the primary experimental observable: In absorption imaging \cite{Foelling2005,Rom2006, Spielman2007}, one directly measures atomic density and its fluctuations. In that case, the strongest correlation features are typically autocorrelation peaks at $d=0$, arising from the perfect correlation of every image with itself. In our case, the fundamental experimental observables are the coordinates of individual particles, from which we derive the correlation functions. In the construction of the correlators, we do not ``double-count" particles, such that all correlators in the main text are free of autocorrelation peaks at zero interparticle distance \cite{FoellingBook2015}. The reported correlators thus directly probe the normal-ordered form of density-density correlators, $\nknknorm =  \langle  \hat{\Psi}^\dagger(k_1) \hat{\Psi}^\dagger(k_2) \hat{\Psi}(k_2) \hat{\Psi}(k_1)\rangle$. Standard results such as Wick's theorem \cite{Wick1950} can readily be applied to the normal-ordered correlation functions and significantly simplify the analysis.

The standard and normal-ordered form of the correlator only differ at the origin and can be related to each other via $\nknk = \nknknorm + \delta(k_1,k_2) \dn{1}$. The normal-ordered correlator is smooth at $d=0$, such that for perfect fermionic antisymmetry, one obtains ${\nknknorm =0}$ for $k_1 = k_2$, consistent with Eq. (2) of the main text.

\subsubsection{Second-Order Density Correlation Function}

Let us consider the second-order momentum correlator expected for an array of $N$ single-fermion sources. For the remainder of the document, normal ordering of correlation functions is implied everywhere.

We define the fermionic field operators $\hat{\Psi} (x)$, in terms of the fermionic annihilation operators $\hat{a}_i$ via
\begin{equation}
    \hat{\Psi}(x) =\sum_{i=1}^N \Phi(x-x_i)\hat{a}_i.
\end{equation}
$\Phi(x-x_i)$ is the on-site wave function at lattice position $x_i$, with $\tilde{\Phi}(k)$ its Fourier transform.

The momentum representation of the field operator is then 
\begin{equation}
 \hat{\Psi}(k)=\sum_i \tilde{\Phi}(k)e^{-i k x_i} \hat{a}_i.
\end{equation}

The momentum density is given by 
\begin{align}
  \langle \hat{n}_{k_1} \rangle & =  \langle  \hat{\Psi}^\dagger(k_1) \hat{\Psi}(k_1) \rangle \lend
   & =  \lvert \tilde{\Phi}(k_1) \rvert^2 \sum_{j,k} e^{i k_1(x_j-x_k)} \langle \hat{a}^\dagger_j \hat{a}_k\rangle \lend
   & = N  \lvert \tilde{\Phi}(k_1) \rvert^2.
      \label{eqn:density}
\end{align}

The expectation value $\langle \hat{a}^\dagger_j \hat{a}_k \rangle$ has to be evaluated with respect to the product state containing one particle per mode, which gives $\langle \hat{a}^\dagger_j \hat{a}_k \rangle=\delta_{jk}$.

The two-point correlation function is given by
\begin{align}
 \nknk & =  \langle  \hat{\Psi}^\dagger(k_1) \hat{\Psi}^\dagger(k_2) \hat{\Psi}(k_2) \hat{\Psi}(k_1)\rangle \lend
   & =  \lvert \tilde{\Phi}(k_1) \rvert^2 \lvert \tilde{\Phi}(k_2) \rvert^2 \lend 
   & \, \sum_{j,k,l,m} e^{i [k_1(x_j-x_m)+k_2(x_k-x_l)]} \langle \hat{a}^\dagger_j \hat{a}^\dagger_k \hat{a}_l \hat{a}_m \rangle.
\end{align}

 Due to fermionic statistics, the expectation value takes the form $\langle \hat{a}^\dagger_j \hat{a}^\dagger_k \hat{a}_l \hat{a}_m \rangle = \delta_{jm} \delta_{kl} -  \delta_{jl} \delta_{km}$. The correlator hence is 
\begin{align}
 \frac{\nknk}{ \lvert \tilde{\Phi}(k_1) \rvert^2 \lvert \tilde{\Phi}(k_2) \rvert^2} & = \sum_{j,k}\left(1-e^{i (k_1-k_2) (x_j - x_k)}\right ) \lend
 & = \sum_{<j,k>}\left(2-2\cos{ (d (x_j - x_k)}\right ) \lend
 \label{eqn:twopoint}
\end{align}

Summation over distinct pairs of sites is denoted by $\langle j,k\rangle$ and $d=k_1-k_2$. Together with Eq.\,(\ref{eqn:density}), this result gives $\ctwo$ in Eq.\,(2) of the main text.

\subsubsection{Wick's Theorem}

The correlations of non-interacting fermionic states can be factorized into lower-order correlators using Wick's theorem \cite{Wick1950}.
\begin{equation}
\nknk = \dn{1} \dn{2} -G^{(1)}_{k_1,k_2}G^{(1)}_{k_2,k_1}
\end{equation}
Here, $ \langle \hat{n}_{k_1} \rangle = \langle  \hat{\Psi}^\dagger(k_1) \hat{\Psi}(k_1) \rangle$ and $G^{(1)}_{k_1,k_2} = \langle  \hat{\Psi}^\dagger(k_1) \hat{\Psi}(k_2) \rangle$ denote the diagonal and off-diagonal part of the one-body density matrix, respectively. 
Also $G^{(1)}_{k_1,k_2} = G^{(1)*}_{k_2,k_1}$ such that $G^{(1)}_{k_1,k_2}G^{(1)}_{k_2,k_1}=|G^{(1)}_{k_1,k_2}|^2$.
For the third order correlations
\begin{align}
\langle \hkn{1} \hkn{2} \hkn{3} \rangle & = && \dn{1}\dn{2}\dn{3} -\dn{1} G^{(1)}_{k_2,k_3} G^{(1)}_{k_3,k_2} \nonumber\\
							 & &&-\dn{2} G^{(1)}_{k_1,k_3} G^{(1)}_{k_3,k_1}-\dn{3} G^{(1)}_{k_1,k_2} G^{(1)}_{k_2,k_1}\nonumber\\
							 & &&+G^{(1)}_{k_1,k_2}G^{(1)}_{k_2,k_3}G^{(1)}_{k_3,k_1} \lend
							 & &&+G^{(1)}_{k_1,k_3}G^{(1)}_{k_3,k_2}G^{(1)}_{k_2,k_1} \nonumber\\ 
							 & =&& \dn{1}\dn{2}\dn{3}-\dn{1}|G^{(1)}_{k_2,k_3}|^2\nonumber\\
							 & &&-\dn{2}|G^{(1)}_{k_1,k_3}|^2 -\dn{3}|G^{(1)}_{k_1,k_2}|^2 \nonumber\\
						    	 & &&+G^{(1)}_{k_1,k_2}G^{(1)}_{k_2,k_3}G^{(1)}_{k_3,k_1}\lend
							 & &&+G^{(1)}_{k_1,k_3}G^{(1)}_{k_3,k_2}G^{(1)}_{k_2,k_1} \nonumber\\ 
							 &= && -2 \dn{1}\dn{2}\dn{3}\ \nonumber\\
							 & &&+ \dn{1}\langle  \hkn{2}\hkn{3}\rangle+\dn{2}\langle \hkn{1}\hkn{3} \rangle \lend
							 & &&+\dn{3}\langle \hkn{1}\hkn{2}  \rangle\nonumber\\
							 & &&+2 \Re\{G^{(1)}_{k_1,k_2}G^{(1)}_{k_2,k_3}G^{(1)}_{k_3,k_1}\}
							\end{align}

The most common form (denoted by the horizontal bar) for the disconnected correlation function at third order is \cite{Cantrell1970,Schweigler2017}
\begin{align}
\overline{\nknknk}_\textrm{dis} =   &-2 \dn{1}\dn{2}\dn{3}\nonumber \nonumber\\
							 &+ \dn{1}\langle  \hat{n}_{k_2}\hat{n}_{k_3} \rangle+\dn{2}\langle \hat{n}_{k_1}\hat{n}_{k_3} \rangle \lend
							 &+\dn{3}\langle \hat{n}_{k_1}\hat{n}_{k_2} \rangle
							 \label{StandardConnected}
\end{align}
With this definition, one finds 
\begin{equation}
\overline{\nknknk}_\textrm{con} =  2 \Re\{G^{(1)}_{k_1,k_2}G^{(1)}_{k_2,k_3}G^{(1)}_{k_3,k_1}\}.
\end{equation}

It is precisely the triple propagator that gives rise to intrinsic correlations at third order \cite{Liu2016}. 
From measurements at first and second order, we can extract the magnitudes $|G^{(1)}_{k_1,k_2}|$ of the propagators, but not their phase. The full information about the propagator $G^{(1)}_{k_1,k_2}$ could be obtained by performing an additional first order interference experiment. With this information, a full factorization of all correlation functions could be achieved. 

\subsubsection{Particle Number Conservation}
We note that in states with fixed particle number, significant correlations can arise purely due to the conservation of total particle number. Intuitively speaking, detecting one of $N$ completely uncorrelated particles at a particular location reduces the probability of finding a particle at a different location by $1-\frac{1}{N}$. This correlation term is absent in the case of pure fermionic statistics on account of the Pauli principle. However, any degree of distinguishability introduced, most notably by integrating out transversal excitation in the tweezer in our effective one-dimensional description, will reintroduce correlation terms due to particle number conservation. In the thermodynamic limit $N\rightarrow \infty$, such correlations can be neglected, but they are highly significant for few-body states. 

We therefore modify the standard definition of disconnected and intrinsic correlation functions to include the scaling factors $s_1(N)$ and $s_2(N)$ given in Eq.\,(4) of the main text. This renormalization of terms in the correlation function ensures that: (a) Correlations from number conservation are removed, i.e. uncorrelated particles give $\cthree_\textrm{con}=0$ for any $N$ and (b) in the limit $N\rightarrow \infty$ our definition of the disconnected correlation functions agrees with the standard form from Wick's theorem (Eq.\,(\ref{StandardConnected})). The baseline of intrinsic correlations, $\cthree_\textrm{con}=0$, then refers to the expectation for exactly $N$ uncorrelated particles.

\subsubsection{Explicit Correlation Functions}
Using the framework outlined above, we explicitly calculate the expected normal-ordered correlator for three fermions emitted from three sources spaced by $a_{12}$ and $a_{23}$.
The correlators for $N$ particles are normalized such that

\begin{align}
	\int  \langle \hat{n}_{k_1} \rangle \diff k_1 & =N \lend
	\iint  \langle \hat{n}_{k_1} \hat{n}_{k_2} \rangle \diff k_1 \diff k_2 &=N(N-1) \lend
    	\iiint \langle \hat{n}_{k_1} \hat{n}_{k_2} \hat{n}_{k_3} \rangle \diff k_1 \diff k_2 \diff k_3 &=  N(N-1)(N-2).
\end{align}

Measuring momentum in units of the lattice momentum $\klattice=\pi/a_{12}$, we obtain for $a_{23}=a_{12}$:
\begin{align}
\langle \hat{n}_{k_1}\hat{n}_{k_2} \rangle = \frac{\dn{1} \dn{2}}{2}(3&-2\cos(2 \pi (k_1-k_2)) \lend
 &- \cos(4 \pi (k_1-k_2)))
\end{align}
and for $a_{23}=1.5a_{12}$
\begin{align}
\langle \hat{n}_{k_1}\hat{n}_{k_2} \rangle = \frac{\dn{1} \dn{2}}{2}(3&-\cos(2 \pi (k_1-k_2)) \lend
 &- \cos(3 \pi (k_1-k_2)) \lend
 &- \cos(5 \pi (k_1-k_2))).
\end{align}
The theoretically expected correlation functions $\ctwo$ are plotted with full contrast in the lower panel of Fig.\,3(a) of the main text.
For the third-order correlator we obtain for $a_{23}=a_{12}$
\begin{align}
\langle  \hat{n}_{k_1}\hat{n}_{k_2} \hat{n}_{k_3}\rangle= & \frac{2 \dn{1} \dn{2} \dn{3}}{27}(3 \lend
&-2\cos(2 \pi(k_1-k_2)) - \cos(4\pi(k_1-k_2))\nonumber\\
&-2\cos(2 \pi(k_1-k_3)) - \cos(4\pi(k_1-k_3))\nonumber\\
&-2\cos(2 \pi(k_2-k_3)) - \cos(4\pi(k_2-k_3))\nonumber\\
&+ 2\cos(2\pi(2k_1-k_2 - k_3))\nonumber\\
&+ 2\cos(2\pi(2k_2-k_1 - k_3))\nonumber\\
&+ 2\cos(2\pi(2k_3-k_1 - k_2)))\nonumber\\
\label{eqn15}
\end{align}
and for $a_{23}=1.5 a_{12}$
\begin{align}
\langle  \hat{n}_{k_1}\hat{n}_{k_2} \hat{n}_{k_3}\rangle= & \frac{2 \dn{1} \dn{2} \dn{3}}{27}(3 \lend
&-\cos(2 \pi(k_1-k_2)) - \cos(3\pi(k_1-k_2)) \lend
&- \cos(5\pi(k_1-k_2))\lend
&-\cos(2 \pi(k_1-k_3)) - \cos(3\pi(k_1-k_3)) \lend
&- \cos(5\pi(k_1-k_3))\lend
&-\cos(2 \pi(k_2-k_3)) - \cos(3\pi(k_2-k_3)) \lend
&- \cos(5\pi(k_2-k_3))\lend
&+ \cos(2\pi(3k_1+2k_2 - 5k_3))\lend
&+ \cos(2\pi(2k_1+3k_2 - 5k_3))\lend
&+ \cos(2\pi(-5k_1+2k_2 + 3k_3))\lend
&+ \cos(2\pi(-5k_1+3k_2 + 2k_3))\lend
&+ \cos(2\pi(3k_1-5k_2 + 2k_3))\lend
&+ \cos(2\pi(2k_1-5k_2 +3k_3)).
\label{eqn16}
\end{align}

Figure\,\ref{fig4} shows the theoretical expectation for the correlator shown in Fig.\,4 of the main text.

\begin{figure*}[ht]
	\centering
	\includegraphics{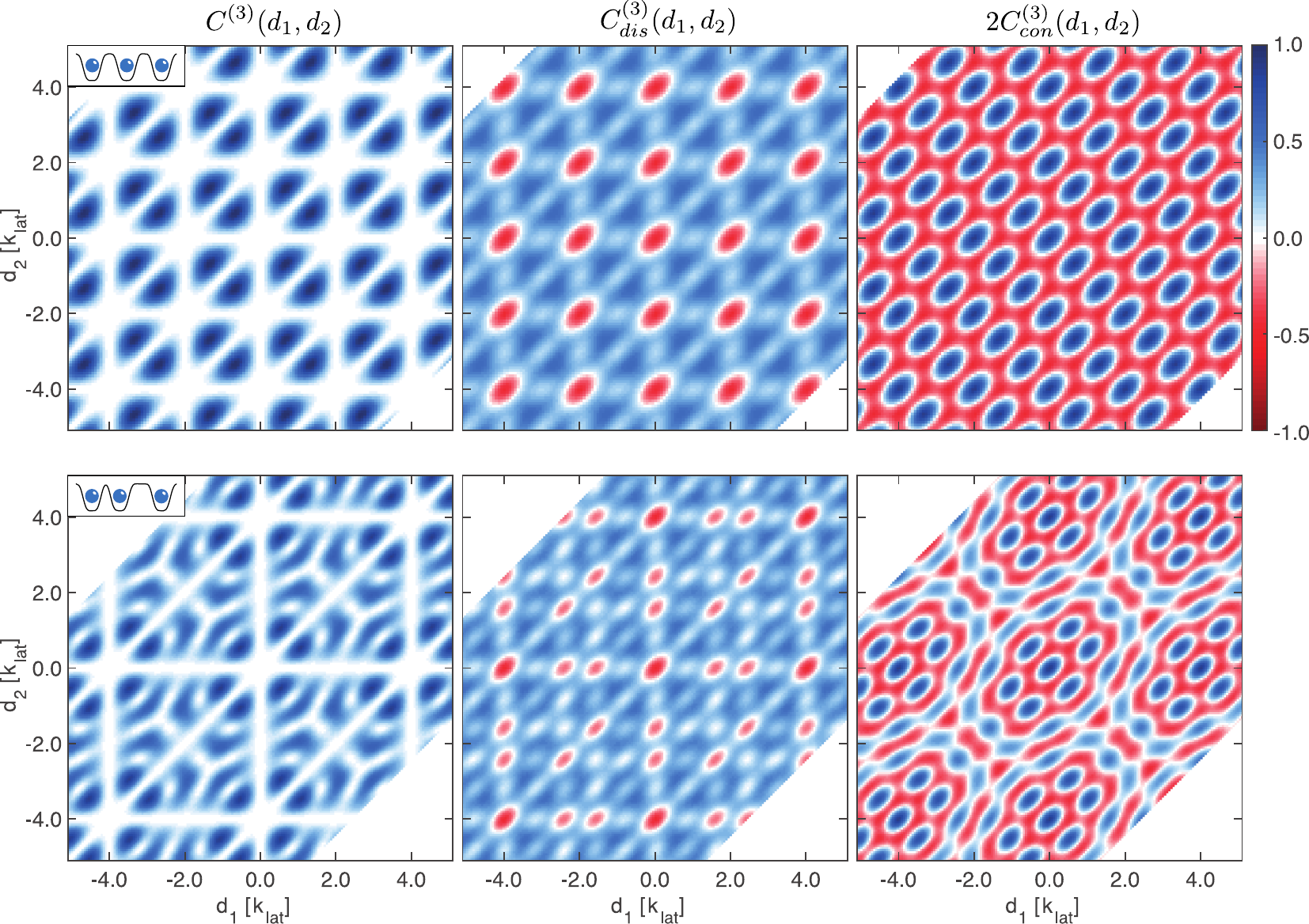}
	\caption{Theoretical expectation for the correlator shown in Fig.\,4 of the main text according to Eqs.\,(\ref{eqn15}) and (\ref{eqn16}).}
	\label{fig4}
\end{figure*}

\end{document}